\documentclass{article}
\pdfoutput=1
\usepackage{arxiv}%

\usepackage{url}

\begin{document}




\title{MEDFORD: A human and machine readable metadata markup language}

  \author{
    Polina Shpilker \\
    Department of Computer Science\\
    Tufts University\\
    Medford, MA\\
    \And
    John Freeman \\
    Department of Computer Science\\
    Tufts University\\
    Medford, MA\\
    \And
    Hailey McKelvie \\
    Department of Computer Science\\
    Tufts University\\
    Medford, MA\\
    \And
    Jill Ashey \\
    Department of Biological Sciences\\
    University of Rhode Island\\
    Kingston, RI\\
    \And
    Jay-Miguel Fonticella \\
    Department of Computer Science\\
    Tufts University\\
    Medford, MA\\
    \And
    Hollie Putnam \\
    Department of Biological Sciences\\
    University of Rhode Island\\
    Kingston, RI\\
    \And
    Jane Greenberg \\
    College of Computing and Informatics\\
    Drexel University\\
    Philadelphia, PA\\
    \And
    Lenore J. Cowen \\
    Department of Computer Science\\
    Tufts University\\
    Medford, MA\\
    \texttt{cowen@cs.tufts.edu}\\
    \And
    Alva Couch \\
    Department of Computer Science\\
    Tufts University\\
    Medford, MA\\
    \texttt{alva.couch@tufts.edu}\\
    \AND
    Noah M. Daniels \\
    Department of Computer Science and Statistics\\
    University of Rhode Island\\
    Kingston, RI\\
    \texttt{noah\_daniels@uri.edu} \\
}




\maketitle

\begin{abstract}
Reproducibility of research is essential for science. However, in the way modern computational biology research is done, it is easy to lose track of small, but extremely critical, details. Key details, such as the specific version of a software used or iteration of a genome can easily be lost in the shuffle, or perhaps not noted at all. Much work is being done on the database and storage side of things, ensuring that there exists a space to store experiment-specific details, but current mechanisms for recording details are cumbersome for scientists to use. We propose a new metadata description language, named MEDFORD, in which scientists can record all details relevant to their research. Human-readable, easily-editable, and templatable, MEDFORD serves as a collection point for all notes that a researcher could find relevant to their research, be it for internal use or for future replication. MEDFORD has been applied to coral research, documenting research from RNA-seq analyses to photo collections.
\end{abstract}

\section{Introduction}

Corals comprise thousands of different organisms, including the animal host and single celled dinoflagellate algae, bacteria, viruses, and fungi that coexist as a holobiont, or metaorganism \cite{bosch2011metaorganisms}. Thus, corals are like cities, rather than the individual animals that inhabit or visit them, as corals provide factories, housing, restaurants, nurseries, and more for an entire ecosystem. Research on coral reefs is ever more pressing, given their local and global contributions to marine biodiversity, coastal protection, and economics and their sensitivity to climate change \cite{hughes2017coral,woodhead2019coral}. Research in this area requires integration of interdisciplinary data across multiple environments and a range of data types: 'omic data such as gene expression data generated using RNA-Seq (RNA transcript sequencing), image and time-lapse video, and physical and environmental measurements including light and water temperature, to name but a few. 
The coral research community has long been committed to sharing and open data formats, and both individual researchers and large funding agencies have invested heavily in making data available~\cite{madin2016trait,liew2016reefgenomics,yu2020sager,donner2017new} and FAIR (findable, accessible, interoperable, and reusable)~\cite{wilkinson2016fair}.



Effective data sharing for coral research, as in all data-intensive domains, requires metadata, which is essential for data organization, discovery, access, use, reuse, interoperability, and overall management~\cite{DBLP:journals/corr/abs-2006-08589}. The growing amount of digital data over the last several decades has resulted in a proliferation of metadata standards supporting these functions~\cite{ball2016rda,qin2012functional}.  However, the proposed mechanisms to create metadata have been focused primarily on the ease of machine parsing and have recommended schema that are cumbersome and difficult for humans attempting to create, edit, or read the metadata. If creating metadata in the appropriate format is difficult, or requires expert curators, then fewer scientists will be able to comply with metadata recommended standards, leading to  scientific data that is not discoverable, and thus not reusable. Meanwhile, an increasing amount of scientific data in multiple countries (including in the US and the EU) now falls under  mandated data sharing policies that  require specification of adequate metadata for discovery. Thus, there is a need for a format that streamlines the process of providing what is mandated by law and policy.  

For the purposes of maintaining and transferring data itself, there already exists a format known as BagIt~\cite{kunze2018bagit} that can handle stable transfer of arbitrary files and their directory structure. BagIt specifies the structure of a zip file that contains an arbitrary directory structure, a payload manifest, and a remote manifest. This  structure upholds the organization of data folders, ensuring that related files can remain within the same subfolder. The payload manifest ensures that all data is transferred without error by storing the data's hash prior to data transfer. The remote manifest also allows researchers to specify remote files that are relevant to the data in the bag. However, the BagIt structure has no inherent descriptor of metadata, although it acts as a convenient means of theoretically transferring metadata. Essenially, BagIt has file-based metadata, but does not proscribe a specific metadata format.


Our research team is building on top of BagIt by developing and implementing the MEtaData Format for Open Reef  Data (MEDFORD). The MEDFORD markup language file format is simultaneously human and machine writable and readable. 
In this regard, we are inspired by the  specification language for the Protein Data Bank (PDB)~\cite{berman2000pdb, young2018worldwide}. PDB files are easily machine-parsable, but unlike JSON files or other commonly-used database submission file types, are also easily human-readable. This human-readability allows for human verification of their contents, although PDB files are still too complex for manual writing. Unlike the PDB format, MEDFORD is intended to be extensible.
MEDFORD is designed to work in conjunction with BagIt's filesystem convention, allowing easily accessible and interoperable bundles of data and metadata to be created and stored. The MEDFORD language is currently implemented as the \texttt{medford} parser, which is itself written in Python.

MEDFORD is initially targeted at coral holobiont transcriptomics data  and coral image collections, with the subsequent goal of supporting metadata for additional research fields. The urgent need for international collaboration around saving coral reefs, plus the sheer complexity of the types and modalities of data the coral scientific community generates (from omics data, to image data with geospatial and temporal components, to temperature and color measurements), make corals a good domain choice. This paper provides the rationale for current work and introduces the MEDFORD (version 1.0) metadata scheme.

MEDFORD will enable interdisciplinary coral reef data to be FAIR~\cite{wilkinson2016fair}. We are currently building the back-end infrastructure to translate between MEDFORD and make it compatible with other existing databases and systems such as Resource Description Framework (RDF), ultimately supporting the interoperability and reusability in FAIR as well; export to RDF is planned for version 1.1 of the \texttt{medford} parser.

This paper reports on our first use case, which focuses on the coral holobiont. Specifically, we focus on  coral holobiont transcriptomics data  (e.g., RNA-Seq, one of the most powerful and common types of omics experiments to explore the genetic basis of factors that lead to coral resilience or vulnerability to environmental stressors), where we build the needed complexity to manage spatial-temporal holobiont expression metadata into MEDFORD from the start. We chose this use case due to difficulty we experienced collecting and organizing metadata about existing coral transcriptomics datasets. A coral researcher, untrained in programming and not a database expert, will be able to directly produce and interpret MEDFORD files more easily than working with RDF authoring tools. We have developed the \texttt{medford} parser to automatically translate MEDFORD files into existing file format standards for depositing in databases and repositories. MEDFORD will enable transcriptomic data to be findable, accessible, and interoperable. While MEDFORD is capable of becoming a general-purpose metadata format, we are implementing the specific use case of coral data to both provide a proof of concept and to aid the coral research community via a set of detailed metadata constructions specific to coral research. 
An extended abstract describing MEDFORD appeared as ~\cite{10.1007/978-3-030-98876-0_18}.

\section{MEDFORD Design Principles}\label{design}

Languages proposed for metadata specification normally consider ease of \textit{either} human generation and parsing, or machine generation and parsing. Human-legible formats, such as unstructured text files, are easy to write but difficult to store in databases or even provide publicly. Meanwhile, highly structured formats such as RDF and JSON are exceptional for import into databases, but are nearly impossible for a researcher to write on the fly. MEDFORD fills a previously unmet need by intentionally balancing ease of human and machine generation and parsing simultaneously.

In addition to being designed as both a machine and human readable and writable format, we also decided that MEDFORD describes the entirety of a project's metadata within a single file. This allows MEDFORD to be extremely lightweight, and it can be simply incorporated into a BagIt bag without any modification. This ensures that if a MEDFORD file is created, it is straightforward to transfer it alongside its related data.

MEDFORD's design principles are informed by the those underlying highly successful metadata standards, such as the Dublin Core ~\cite{weibel2000dublin}, Ecological Metadata Language (EML)~\cite{fegraus2005maximizing}, and the Data Document Initiative (DDI)~\cite{vardigan2014ddi}, while addressing additional requirements enabling ease of metadata creation and other aspects. The design requirements for creating MEDFORD are as follows:
\begin{enumerate}
\item A mechanism for use by scientists at the point of data collection. 
\item A human-readable and human-understandable format for specifying metadata. 
\item A simple and easily understandable syntax for specifying metadata elements.
\item The ability to create and reuse templates for specifying metadata for common data types. 
\item Applicability beyond the coral use case, to other research domains.
\item The ability to author metadata in a user's preferred text editor without a dependency on special-purpose software. 
\item The ability to detect and explain errors in metadata specification via easily understandable error messages targeted toward scientists. 
\item Automatic translation into a number of useful machine-readable formats after initial specification, including the Resource Description Format (RDF), Extensible Markup Language (XML) and JavaScript Object notation (JSON), as well as database formats. This could ease some costs currently incurred; for instance, BCO-DMO removes the responsibility for translating to different formats from the scientist to that of a specialized curator, at substantive cost. 
\end{enumerate} 

These requirements are justified by past experience of scientists crafting metadata~\cite{qin2012functional}. Web-based metadata interfaces can be cumbersome when one is entering metadata for a set of similar data files or publications. The machine-readable formats XML, RDF, and JSON are difficult to understand and edit for scientists who are not programmers. Furthermore, error messages for mistakes in XML, RDF, and JSON specifications are cryptic for those same people. Plain-text specification of metadata, such as the NSF's BCO-DMO resource~\cite{chandler2016bco} is intensive in human labor for those who must then translate it into a machine-readable form in order for BCO-DMO to ingest such data (currently, BCO-DMO requires metadata to be submitted in Rich Text Format, and it is then transcribed by a human operator). Thus, there is a need for an intermediate format that is both machine-readable and human-readable and understandable by the scientists most qualified to specify the metadata correctly. 

MEDFORD aims to solve problems associated with specifying interdisciplinary research metadata, as demonstrated by our initial use case applied to coral reef 'omics data. Coral researchers study a wide variety of properties of corals (bioinformatics, growth, bleaching, phylogeny) and for a variety of purposes (ecology, basic biology, biomedicine). Connecting the work of this diverse group of researchers requires developing sustainable scientific databases so that researchers can discover each others' datasets, integrate them into more novel research, and support further scientific discovery. These databases need to support both accurate analysis of research as well as data discovery and reuse. In general, however, the principles above apply to any scientific metadata specification problem, and the specific extensions identified here may be supplemented for other scientific disciplines. Thus MEDFORD can be used as a tool for metadata creation in any scientific discipline.

The requirements above are realized by MEDFORD by adding design elements that satisfy the above principles: 
\begin{enumerate}
\item A contextual grammar, devoid of parentheses and the need to close clauses with specific end statements. 
\item A simple way to denote kinds of metadata, starting with an \texttt{@}, and containing at most three parts: the major tag, minor tag, and the metadata itself. A major tag (such as \texttt{@Contributor}) indicates the type of metadata being described, while a minor tag (such as \texttt{ORCID} in the context of \texttt{@Contributor-ORCID}) indicates the name of the metadata attribute being described.
\item A two-level hierarchy based on major and minor tags organizes the metadata into categories and subcategories which provide the relational structure without compromising the simplicity of the metadata description.
\item A simple concept of user-extensible formatting, in which metadata details not covered by the main keywords can be added via notes.
\end{enumerate} 

Consider the following example of a \texttt{@Contributor} clause, where \texttt{@Contributor} is the major tag and \texttt{ORCID} and \texttt{Role} are the minor tags which associate those metadata with that contributor.

\begin{verbatim}
@Contributor Hollie M. Putnam
@Contributor-ORCID 0000-0003-2322-3269
@Contributor-Role Corresponding Author
\end{verbatim}

If we wanted to additionally include this contributor's email address, we simply add an additional line:

\begin{verbatim}
@Contributor Hollie M. Putnam
@Contributor-ORCID 0000-0003-2322-3269
@Contributor-Role Corresponding Author
@Contributor-Email hputnam@uri.edu
\end{verbatim}

\subsection{The MEDFORD Language Syntax}\label{sec2}

In this section we discuss the principles of the design of the syntax of a MEDFORD file format for metadata. MEDFORD is written in UTF-8 though all reserved tokens and characters fall within the ASCII range, while user-defined tags may use extended UTF-8 characters. MEDFORD tags are indicated with the \texttt{@} character.  
Anything after an \texttt{@} character, until the next space in the file, is read as a tag by the \texttt{medford} parser. 
There are two other protected symbols that have special meanings in the MEDFORD language: these are \texttt{\#} which is treated as a comment character: characters
after a \texttt{\#} on the same line are ignored and not processed by the \texttt{medford} parser. Finally, the \texttt{\$\$} string (two dollar signs in a row) is used to indicate the beginning and end of a string
that should be parsed by \LaTeX math mode: this enables a MEDFORD language parser to either render or pass through special characters from raw MEDFORD files, in which non-ascii characters are strongly discouraged. 


The following design  principles are important in MEDFORD file syntax: 
\begin{itemize}
    \item MEDFORD files use the ASCII character set whenever possible. The characters \texttt{@}, \texttt{\#}, and \texttt{\$\$} (as an enclosing pair to denote LaTeX source) are reserved and protected. 
    \item MEDFORD tags are referred to as \texttt{@}-tags and always start with the \texttt{@} character. Particular \texttt{@}-tags are given meanings, and formatting requirements and rules that are either recommended or required
    \item If a version of the MEDFORD language parser encounters an \texttt{@}-tag it does not recognize, the parser passes its associated text  through verbatim, treating it identically to how \texttt{@COMMENT} is treated. Thus, scientists are free to make up new tags that extend what is currently defined in the language.
    \item MEDFORD is initially being developed for corals data, and so \texttt{@Date} and \texttt{@time} and geospatial coordinate data are common and important. These tags have recommended standardized ASCII formats, and the \texttt{medford} parser does type checking on these fields. These tags have corresponding \texttt{*-Unstructured} equivalents for flexibility, which are not type checked. For instance, \texttt{@Date-Unstructured} might be used to denote part of a date, where the precise date is unknown (for instance, ``Fall 2021'').
\end{itemize}


To make MEDFORD files more easily human-readable, considering our analogy to the protein databank, we adopted a similar approach to the Fasta file format, where each header line is distinguished by an `\texttt{>}' symbol. We chose to use the `\texttt{@}' character, as it is commonly used for tagging users or keywords in systems like GitHub and Twitter, and is not found in everyday text except for emails. Therefore, a line headed by an `\texttt{@}' symbol can be assumed to be a MEDFORD tag in all cases. Meanwhile, later `\texttt{@}' characters have no effect on the \texttt{medford} parser, as only `\texttt{@}' characters at the very beginning of a line matter.


This is best explained by example. Consider the example of specifying a pipeline used for RNA-seq analysis of coral data (note that the metadata associated with a tag can be arbitrarily long and may span multiple lines):

\begin{verbatim}
@Software R
@Software-Version 4.0.4 ("Lost Library Book")
@Software-Notes Packages used include dplyr, stringr,
    and genefilter.

@Software DESeq2
@Software-Version 1.28.1
@Software-Notes Used as a package in R.
@Software-Notes Installed through BioCManager.
\end{verbatim}

The {\tt Software} tag specifies a piece of software involved in the research. In this case, R is being described as a relevant piece of software, with a {\tt Version} tag used to specify what version of R was used as this is critical information. An important feature of this example is the arbitrary difference between the way R packages are described. The author has determined that DESeq2 is a critical package, so decided to use a separate {\tt Software} tag to describe it. Meanwhile, dplyr and stringr were useful in the analysis but not critical, so were left as {\tt Notes} on the R {\tt Software} block. This showcases one of the strengths of the MEDFORD file format; researchers are free to determine whether something is important enough to warrant having a dedicated {\tt Software} tag, or if they can be listed as an arbitrary {\tt Note} on a parent piece of software.

\subsection{MEDFORD Data Provenance}

One of the major goals of MEDFORD is to enable the simple association and description of related but possibly separate data resources. The BagIt filesystem convention~\cite{kunze2018bagit} provides a convenient way to wrap multiple files into a consistent directory structure. However, BagIt's own metadata capabilities are limited to describing the files present or how to fetch them from a network. By including a MEDFORD file into a bag, we are able to therefore describe the metadata as well as reference or include the data themselves. MEDFORD does not try to supplant the W3C data provenance standard (RDF) but rather provide a tangible, simple format that meets users' need. A MEDFORD metadata description could be automatically converted into an RDF representation.

All MEDFORD files are defined in reference to a BagIt bag, although the special use-case of an empty bag is common and acceptable. The BagIt bag binds a set of files to the MEDFORD file according to the BagIt standard, where these files describe a variety of resources, including source code, scientific papers, or raw data, each represented by a major tag in the MEDFORD file. The versioning and origins of that file are marked using a secondary major tag, where the tag can represent that the bag is considered to be the primary and authoritative source for the data or resource. Other secondary major tags describe the file as either a copy of an existing source, or simply a pointer to a URI where the resource can be obtained. 

\vspace{0.5cm}

{\small
\begin{tabular}{|c|c|c|} \hline
    \texttt{@Data\_Primary} & \texttt{@Code\_Primary} & \texttt{@Paper\_Primary} \\
    \texttt{@Data\_Copy} & \texttt{@Code\_Copy} & \texttt{@Paper\_Copy} \\
  \texttt{@Data\_Ref} & \texttt{@Code\_Ref} & \texttt{@Paper\_Ref} \\ \hline
  \end{tabular}
} 

\vspace{0.5cm}

MEDFORD's place in a BagIt directory structure is that the MEDFORD (\texttt{.mfd}) file is placed at the top level of the bagit directory structure. Any files carried along in the BagIt archive exist as \texttt{Copy} or \texttt{Original} directives (whether \texttt{Data}, \texttt{Code}, or \texttt{Paper}). The BagIt \texttt{manifest-sha512.txt} manifest refers to these files and their checksums. In contrast, any files only referred to using \texttt{Ref} directives are not listed in the BagIt manifest and are instead described in the BagIt's \texttt{fetch.txt} as remote resources.

\texttt{@Data\_Primary} and \texttt{@Data\_Copy} both refer to resources that have been packaged with the MEDFORD metadata file, and should be available from the bag in a self-contained fashion, without having to visit external sources. From the point of view of the bag itself, there is no difference between these two tags; the difference is based on user context: \texttt{@Data\_Primary} means that the BagIt bag is considered to be the primary and authoritative source for the data or resource; \texttt{@Data\_Copy} means that BagIt has placed a copy of the data or resource into the bag, but that it does not claim the primary role. Finally \texttt{@Data\_Ref} refers to DOIs, URLs, or other pointers to data or resources that are {\em not} placed in the bag, but rather represent external databases or resources. 

Here, we provide two potential use cases as examples.

Example use case: Researchers wish to create an index of all publicly available RNAseq raw data that have been released on the Internet. They create a MEDFORD file to point to all these data resources, but they will store none of these themselves; the MEDFORD file will just be an index, and all \texttt{@Data} tags will be of the form \texttt{@Data\_Ref}. This is an example MEDFORD file which would be associated with an empty bag. 

    Example use case: Researchers wish to store all the necessary data and programs necessarily to replicate their RNAseq analysis. They are the owners/collectors of the raw transcriptomic data, which they do downstream analysis using a couple of small home-grown scripts to filter bad reads, but then complete their downstream analysis using several popular software packages, including STAR and DESEQ2. They then used a novel dimension reduction package called SQUISHSEE  from other researchers to visualize their results. They elect to include their transcriptomic data and homegrown scripts in the bag, and use \texttt{@Data\_Primary} and \texttt{@Code\_Primary} tags to reference them. The \texttt{@Code\_Primary} tag is not appropriate for STAR, DESEQ2 and SQUISHSEE, since they do not own or maintain these resources; they need to decide whether to use \texttt{@Code\_Copy} and place a copy of  these resources into the bag or not, in which case they would use instead the \texttt{@Code\_Ref} tag. In this case, because STAR and DESEQ2 are well-maintained and supported standard packages, they elect \texttt{@Code\_Ref}, and don't include a copy of the code in the bag. On the other hand SQUISHSEE is only used by a handful of researchers, and they worry about its longevity. Thus they also put a copy of the version of SQUISHSEE  they are using in the bag, with a \texttt{@Code\_Copy} tag. Later, when DESEQ2's new update uses a library that is not completely standard, they update the bag and decide to put a copy of the old version of DESEQ2 into the bag, just in case. 
    
A specification document for the MEDFORD language is available at \url{https://github.com/TuftsBCB/MEDFORD-Spec}.

\section{Reusability}

\subsection{Tag Extensibility}
MEDFORD has a set of pre-defined major and minor tags that it uses for conversion into various other formats, but if a user cannot find a tag that they believe suits the metadata that they are storing, they can simply define one of their own without any additional overhead. All the user must do is use it as if it were already defined, and the data and its structure will be read by the MEDFORD parser. Any novel tags defined this way will be treated as \texttt{*-Unstructured} tags, and not validated, though they will persist across copies of the MEDFORD file. This provides a dynamic aspect whereby any model  created or adjusted to include the new user-defined tag could be output to any secondary formats without any changes in MEDFORD structure.

\subsection{MEDFORD Templates}

Due to the simple plaintext structure of a MEDFORD file, it is easy to create templates. A MEDFORD file can be partially filled out, saved, copied, and then re-used. For example, a lab may template out a list of contributors and funding sources and when an individual needs to create a MEDFORD file they simply create a copy of this MEDFORD file and change contributor roles as necessary before filling out the rest of the file. MEDFORD files describing similar data may also be re-used like this. 

For example, consider a researcher who commonly works on one species of coral, such as \textit{Pocillopora damicornis}. The researcher could use a MEDFORD template with the commonly-used tags filled in, shown below:

\begin{verbatim}
    @Species Pocillopora damicornis
    @Species-Loc Sabago Isthmus, Panama
    @Species-ReefCollection 06/12/20
    @Species-Cultured University of Miami Coral Resource 
        Facility
    @Species-CultureCollection 06/21/20
\end{verbatim}

For further reuse, the researcher may also include MEDFORD's ``invalid value'' token, which can be used to force users of a template to fill it out with complete information. The \texttt{medford} parser would require the user to fill in the specific placeholders (\texttt{[..]}) prior to validation. This eliminates the possibility that a researcher could accidentally leave a value for an older version of the template, further error-proofing MEDFORD templates. The same template, but using these reserved template tokens, is shown below:

\begin{verbatim}
    @Species Pocillopora damicornis
    @Species-Loc Sabago Isthmus, Panama
    @Species-ReefCollection [..]
    @Species-Cultured University of Miami Coral Resource
        Facility
    @Species-CultureCollection [..]
\end{verbatim}

In future work, we plan to include a \texttt{`\#include} directive which allows the contents of one file to be imported into and validated in the context of another MEDFORD file.

\subsection{MEDFORD Macros}

To further alleviate the workload placed on researchers to document their work, MEDFORD includes the concept of a macro. Similar to a variable defined in BASH, a macro is a string name that is directly replaced with another, longer string. In MEDFORD, a macro is defined by specifying a backtick (\texttt{`}), \texttt{@}, a one-word name, and the macro body (which can contain multiple lines, ending at the next reserved word, which could begin another macro definition or could be a regular tag). For example, in lieu of typing their institute five times to document each of their collaborators at the same institution, they can define a macro as follows:
\texttt{`@myinstitute 100 Institute Drive, State, Zip}

Everywhere \texttt{`@myinstitute} is used, it will automatically be replaced by \texttt{100 Institute Drive, State, Zip}.

\subsection{Backend Extensibility}
Many other formats are simple to add to the MEDFORD parser. For example, one may wish to submit their data to a database such as BCO-DMO~\cite{chandler2016bco}, which requires an RTF (rich text format) file structure with unique content requirements. For example, for a data submission, BCO-DMO requires at least some form of identification for which \texttt{@Expedition} the samples were collected on. This identification, may be either some combination of \texttt{ShipName} or \texttt{CruiseID}, etc. In defining a backend translation to BCO-DMO for MEDFORD, the MEDFORD parser can ensure this is upheld. This ability to act as an intermediary allows for a lab to write a single MEDFORD file to describe their research and export it to a multitude of different formats.

Similarly to BCO-DMO, other formats can be added to the MEDFORD parser easily; tutorials will be available in the github repository. It is worth noting that metadata associated with user-defined tags will not be parsed but simply passed along verbatim. For instance, the \texttt{@Image-Coverage} tag in the coral image data example below does not specify any units; if this tag were expected by some destination format, units might be assumed by convention, but would otherwise be left to the user (the user could also specify units in plain text in the metadata, e.g. ``6.2 degrees'').

\section{Example Medford Files}

\subsection{MEDFORD File for RNA-Seq data} 

\begin{verbatim}
    @Method Illumina HiSeq2500
    @Method-Type Sequencing
    @Method-Company Dovetail Genomics, Santa Cruz, CA, 
        USA
    @Method-Sample Healthy
    @Method-Note Chicago libraries, more sensitive to 
        DNA size 
    
    @Code_Ref HiRise 
    @Code_Ref-Type Assembly of genome scaffolds 
    
    @Code_Ref BLAST
    @Code_Ref-Type Identify and remove scaffolds of 
        non-coral origin 
    @Code_Ref-Note Searched against databases from 
        Symbiodiniaceae, Bacteria, and viruses 
\end{verbatim}

\subsection{MEDFORD File for coral image data} 

\begin{verbatim}
    @Image 05-01-19_Image3
    @Image-Date 2019-05-01T19:20:30.45
    @Image-Site LTER 4
    @Image-Habitat Outer 10m
    @Image-Pole 3-4
    @Image-Quadrant 4
    @Image-Coral Acropora
    @Image-Coverage  6.2
    
    ...
    
    @Taxonomy Cnidaria
    @Taxonomy-Type Phylum
    
    @Taxonomy Anthozoa
    @Taxonomy-Type Class
    @Taxonomy-Parent Cnidaria
    
    ...
    
    @Region LTER 1 polygon including 
        LTER 0 on north shore
    @Region-NorthernCoord -17.47
    @Region-SouthernCoord -17.49
\end{verbatim}

\section{MEDFORD Implementation}

\subsection{MEDFORD Parser}

The MEDFORD parser, known as \texttt{medford}, essentially has two roles. First, it validates the syntax and structure of a provided MEDFORD file as described earlier. Additionally, the parser validates the content of a provided MEDFORD file, such as ensuring date fields are in the correct datetime format. In the future, we plan to support further validation of specific metadata, such as ORCID, geographic coordinates, and grant numbers from various funding agencies.
The purpose of validation is to ensure that the file is written in correct MEDFORD format, including major and minor tags, and that each tag is being applied to some data. For example, a user cannot describe an ORCID without having some Contributor name with which to associate it. We note that the current MEDFORD specification is silent as to whether or not to preserve the ordering of tags (for example \texttt{@Contributor} does not specify an author order and relies on \texttt{@Contributor-Role} to indicate significance). However, the \texttt{medford} implementation will preserve order of tags in a future version.

The second role of the \texttt{medford} parser is to optionally compile an input MEDFORD file into some destination format. The current \texttt{medford} parser specializes in translating a MEDFORD file into a Bag; the \texttt{medford} parser can gather all the files referenced in a given MEDFORD file, and creates a Bag following all BagIt specifications. This Bag can then be used to transfer all of the metadata and data of a research effort. The current plans include to add additional output types in the future, such as RDF.

The \texttt{medford} parser is written in Python (3.8), relying on the Pydantic parsing module to validate the MEDFORD syntax and structure.

Due to the amount of control a compiler has to have over the input, creating a parser normally causes the vocabulary to become extremely defined and controlled. The \texttt{medford} parser, however, was developed specifically to avoid restricting the acceptable vocabulary. While the parser can only validate major and minor tags it is aware of, it will not break on novel inputs. 

As an example, consider the following \texttt{@Code\_Ref} block, which references code from an external source that was referenced in a study. The \texttt{OS} and \texttt{Language} minor tags are novel, and MEDFORD will not perform any validation on them. 

\begin{verbatim}
    @Code_Ref MEDFORD Source Repo
    @Code_Ref-Version 1.0
    @Code_Ref-URI https://github.com/TuftsBCB/medford
    @Code_Ref-Type GitHub
    @Code_Ref-Language Python
    @Code_Ref-OS Linux MacOS
\end{verbatim}

Importantly, the \texttt{medford} parser is specifically developed such that the syntactical parsing logic is entirely separate from the vocabulary definition. Given a desire to begin validating the contents of a novel tag, a user can easily add their own validation without having to interact with the parsing logic. All vocabulary validation definitions are stored entirely independently of the parsing logic, and can be edited with minimal consequences. In this \texttt{@Code\_Ref} example, a user could implement the validation for the \texttt{OS} minor tag to ensure it is some combination of of \texttt{Windows}, \texttt{MacOS}, and \texttt{Linux}.

Given that the \texttt{medford} parser is open-source, a research group may add validation to their local copy of the MEDFORD parser without needing to interact with other groups, though we will be welcoming any and all pull requests to add validation that users feel is missing.

\subsection{Error Handling}

MEDFORD errors come in three major forms:
\begin{itemize}
    \item \textbf{Syntax Errors}: errors in the MEDFORD formatting in the provided file, such as multiple uses of the same macro name.
    \item \textbf{Validation Errors}: errors in the content or format of metadata provided for known major-minor tag combinations. For instance, a \texttt{@Date} field that does not contain a valid datetime string, or a \texttt{@Contributor-ORCID} field whose ORCID is not valid would both constitute validation errors.
    \item \textbf{Missing Data Errors}: Required fields are missing, such as \texttt{@Date} major tag without a corresponding \texttt{@Date-Note} minor tag.
\end{itemize}

All three types of errors are errors that a standard user is expected to encounter during use, especially during first-time use or novel data type description. Special care has been invested in ensuring these errors will be as human-legible as possible.

For all expected errors, \texttt{medford} provides an error text to the user that contains the following information: the line number where the error was encountered, the major-minor tags involved at that line, and an error text description.

Two examples of \texttt{medford}'s error messages and their improvements are shown below. 

First, a standard Pydantic error contains information that is highly specific to the backend implementation of \texttt{medford} parser and irrelevant for standard use.
\begin{verbatim}
Contributor -> 0 -> 1 -> __root__
  Corresponding Authors must have a provided validated email
  (type=value_error.incomplete_data_error)
\end{verbatim}

This has been improved with the addition of the line number in which the error appeared in the \texttt{MEDFORD} file, and removal of the implementation-specific array indices.
\begin{verbatim}
Line 1 : @Contributor has incomplete information: 
  Corresponding Authors must have a provided validated email.
\end{verbatim}

Secondly, some major and minor token combinations may have multiple valid formats, and Pydantic's standard error handling will throw a unique error for each failed format validation. For example, a \texttt{@Date} may currently either be in a Date or DateTime format, as specified by Python:
\begin{verbatim}
Date -> 0 -> 1 -> desc -> 0 -> 1
  invalid date format (type=value_error.date)
Date -> 0 -> 1 -> desc -> 0 -> 1
  invalid datetime format (type=value_error.datetime)
\end{verbatim}

The \texttt{medford} parser automatically consolidates these errors into a singular error for legibility:
\begin{verbatim}
Line 7 : @Date-desc is of the wrong type: 
  invalid date format.
\end{verbatim}

\section{Availability of \texttt{medford} and the Coral RNAseq collection} 
The \texttt{medford} parser is open-source and available under the MIT license at:  \url{https://github.com/TuftsBCB/medford} as well as via \texttt{PyPi} as the package \texttt{medford}, so it can be installed by invoking \texttt{pip install medford}. A specification for the MEDFORD language is available at \url{https://github.com/TuftsBCB/MEDFORD-Spec}.

Some example files are provided in the parser directory, however, a separate repository is also in development for a larger collection of example MEDFORD files. This repository is available on GitHub at: \url{https://github.com/TuftsBCB/MEDFORD-examples}. This repository is a collection of primarily Coral RNA-Seq experiments. This repository also contains partial MEDFORD files for use as templates.

\section{Discussion and Future Work} 
This manuscript presents MEDFORD, a lightweight metadata format initially targeted at coral reef research data, intended to be easy for researchers without programming expertise to create and maintain. Initially supporting the FAIR principles~\cite{wilkinson2016fair} of interoperability and reuse, MEDFORD aims to support all FAIR principles.

Currently, MEDFORD relies on editing ASCII or UTF-8 text, but will soon be able to extract text content from Microsoft Word files.

One possible critique of MEDFORD is the variety of possible tags.
For instance, it may be challenging for a user to remember whether the needed tag is \texttt{@Contributor-Association}, \texttt{@Contributor-Institution}, or \texttt{@Contributor-Location}. A rich template library can mitigate this, by providing examples that a user can simply fill in. A searchable template library portal (similar to \LaTeX's CTAN) would enable users to find applicable templates as the template ecosystem grows. In the future, support for the Language Server Protocol will allow a user of any compatible text editor to get intelligent suggestions and autocompletions for common tags. This will also mitigate the likelihood of minor typographical errors in tags causing them to be unrecognized. To further mitigate the likelihood of typographical errors, the \texttt{medford} implementation will reject a minor token without an accompanying major token. Further user testing and feedback will result in further enhancements to the MEDFORD language and the \texttt{medford} parser implementation.

As a consequence of the MEDFORD parser's compilation use, MEDFORD files have a lifecycle. There are raw, un-validated MEDFORD files, there are validated MEDFORD files, and finally there are MEDFORD files that have been compiled (such as in the case of a BagIt compilation). The difference between these is a critical. A researcher needs to know whether or not a MEDFORD file has been validated before they try to submit it to a database.

A major future goal will be output of RDF and support for linked open data. 
We hope to add the ability to translate a MEDFORD file (and created bag, if applicable) into an RDF, as well as  the data-1 compliance this involves.

An unsolved problem is how to handle multiple authors, and conflicting claims of ownership.
While there is nothing preventing a MEDFORD file from being passed between collaborators, keeping track of changes is a challenge. How can one researcher be certain that they are editting the most recent version of a MEDFORD file? Perhaps even two researchers are editing their own copies of the same MEDFORD file. Technically both are the most up-to-date in their own facet: one researcher added the coral sample metadata while another added the sequencing pipeline metadata. There exist some solutions to this in external tools such as GitHub, but is it viable to ask MEDFORD adopters to use these tools?

One solution we are in the process of considering for a future version of MEDFORD is to implement the concept of the \texttt{include} directive. Rather than restricting a MEDFORD file to a single file, include directives will enable users to work in separate, smaller files that will automatically be combined by the MEDFORD parser. This allows each MEDFORD file to be dedicated to a specific portion of the research project, such as one file for coral sample metadata and another for sequencing pipeline metadata. This partially solves the multiple authorship problem, as each author can be held responsible ensuring all collaborators have the most up-to-date version of the metadata they are authoring.

The ``R'' in MEDFORD currently represents ``reef'' as our initial application domain has been coral reef data. However it stands to reason that the ``R'' might represent ``research'' in the future.

\section{Author contributions statement}

P.S., J.F., L.C., A.C. and N.D. came up with the initial design for MEDFORD; P.S., J.F., H.M., J-M.F, J.A., and H.P. tested and implemented initial MEDFORD examples; P.S., A.C. and N.D. worked on the back end medford parser; P.S., J.G., L.C, A.C., and N.D. helped write and review the manuscript. 
\section{Acknowledgments}
This work is supported in part by funds from the National Science Foundation under NSF grants  OAC-1939263. OAC-1939795 and HDR-BIO NSF-OAC \#1940233.

\section{Competing interests}
The authors declare NO Competing Interest.



\bibliographystyle{plain}
\bibliography{reference}

\end{document}